\documentclass[aps,pra,groupedaddress,showkeys,reprint,floatfix]{revtex4-2}

\usepackage{amsmath}
\usepackage{mathtools}
\usepackage{physics}
\usepackage{xfrac}
\usepackage{txfonts, comment,stmaryrd}
\usepackage[T1]{fontenc}
\usepackage{graphicx}
\usepackage{dcolumn}
\usepackage{bm}
\usepackage{xfrac}
\usepackage{hyperref}
\hypersetup{
colorlinks = true,
linkcolor  = blue,
citecolor  = blue,
urlcolor   = blue
}
\usepackage{mleftright}
\mleftright
\newcommand{\zedex}{z\textsf{-}x}
\newcommand{\exzed}{x\textsf{-}z}

\newcommand{\exex}{x\textsf{-}x}
\newcommand{\exwhy}{x\textsf{-}y}
\newcommand{\whyex}{y\textsf{-}x}
\newcommand{\whywhy}{y\textsf{-}y}
\begin{document}
\title{Strongly Interacting Two-component Coupled Bose Gas in Optical Lattices}
\author{Sagarika Basak}
\email{basak.sagarika@rice.edu}
\author{Han Pu}
\email{hpu@rice.edu\\\\\\} 
\affiliation{Department of Physics and Astronomy \& Rice Center for Quantum Materials, Rice University, Houston, Texas 77251, USA}

\date{\today}

\begin{abstract}
Two-component coupled Bose gas in a 1D optical lattice is examined. In addition to the postulated Mott insulator and superfluid phases, multiple bosonic components manifest spin degrees of freedom. Coupling of the components in the Bose gas leads to substantial changes in the previously observed spin phases, giving rise to new effective spin Hamiltonian and unraveling remarkable spin correlations. The system in the absence of coupling, exhibits ferromagnetic and non-ferromagnetic spin phases for on-site intra-component interaction stronger than inter-component interaction. Upon introduction of coupling, the phase transition switches from first- to second-order. For comparable on-site inter- and intra- component interactions, with coupling, instead of one, two spin phases emerge with a second-order phase transition. Exact diagonalization and Variational Monte Carlo (VMC) with stochastic minimization (SM) on Entangled Plaquette State (EPS) bestow a unique and enhanced perspective into the system beyond the scope of a mean-field treatment.
\end{abstract}


\maketitle

\section{\label{sec:level1}Introduction}
Ultracold gases provide an unparalleled platform to explore physics found in atomic or molecular gases, offering a manifold of prospects to simulate and examine canonical models of strongly correlated electrons in condensed matter systems \cite{Bloch2012,Lewenstein2007}, owing to them being clean, versatile, and highly controllable. Observation of superfluid to Mott insulator transition \cite{Greiner2002} and theoretical and experimental demonstration of exotic quantum phase transitions \cite{Sachdev2000} in cold atomic setups have paved the way towards studying many-body physics \cite{Bloch2012,Lewenstein2007,Bloch2008,Esslinger2010,Weitenberg2011}, in association with strong correlations \cite{Greiner2002,Jaksch1998}, and relating to the emergence of collective and thermodynamic behavior \cite{Morsch2006,Bloch2008}. Experimental demonstrations---of cold atoms in lattices via weak trapping potential superimposed onto optical lattices \cite{Esslinger2010}, achievement of low temperature relevant for spin orders \cite{McKay2011,Bloch2012,CapogrossoSansone2010,Jordens2010} necessary for superexchange couplings resulting in effective nearest-neighbor spin--spin interaction demonstrated in an array of double wells \cite{Flling2007} via a combination of evaporative and adiabatic cooling \cite{Ho2007,Medley2011,Rabl2003,Trebst2006,Kantian2010,Lubasch2011,Sorensen2010,Gammelmark2013}, and simulating magnetic fields via artificial gauge potential \cite{Dalibard2011,Goldman2014}---provide immense relevance to this field. Quantum magnetism, a most fascinating area of research in condensed matter physics, can be reproduced using cold atoms with tunable geometry and parameters; for instance, experimental study of itinerant magnetism in ultracold Fermi systems with repulsive interactions \cite{Jo2009}, classical magnetism in triangular lattices \cite{Struck2011} with fast oscillations of optical lattice enabling tuning the sign of nearest-neighbor tunneling \cite{Eckardt2005}, bosons with strong interactions in a tilted lattice at commensurate fillings for study of Ising model and quantum phase transition \cite{Sachdev2002,Simon2011} and possible realization of quantum dimer models \cite{Sachdev2002,Pielawa2011}.

Bose--Hubbard Hamiltonian describing the dynamics of bosonic atoms is a quintessential model to probe strongly correlated many-body quantum systems \cite{Fisher1989,Lewenstein2012,Bloch2008} and simulate lattice spin models \cite{Aspuru2012,Blatt2012,Bloch2012,Cirac2012,Hauke2012,Houck2012,Lewenstein2012,Lewenstein2007,Gross2017,Kuklov2003,Altman2003} such as quadratic-biquadratic spin model \cite{Ripoll2004}, or antiferromagnetic spin chains \cite{Sachdev2002} and spin-$1$ model exhibiting Haldane (gapped) insulator phase \cite{Torre2006,Berg2008,Amico2010,Dalmonte2011,Rossini2012}. Interesting extensions of Bose--Hubbard model, such as inclusion of next-to-nearest-neighbor tunneling or long range interactions \cite{Trefzger2008, Zhou2010,Jin2013,Mishra2009,Kumar2011,Bai2020}, spinor bosons with multiple internal degrees of freedom \cite{Kurn1998,Kurn2013} and addition of nonlinear coupling between the components \cite{Krutitsky2004,Kimura2005,Zhou2011,Pietraszewicz2012,Li2012,Sowiski2013}, display remarkably rich phase diagrams \cite{Zhou2010,Zhou2011,Pietraszewicz2012,Li2012,Sowiski2013}. 
A two-component Bose--Hubbard in Mott phase reveals pseudospins effectively coupled by Heisenberg exchange \cite{Kuklov2003} imitating spin-$\sfrac{1}{2}$ Hamiltonian ideal for study of quantum magnetism \cite{Kuklov2003,Duan2003,Giuliano2013}, displays a spin-Mott phase in spin-$1$ Hamiltonian \cite{CapogrossoSansone2010,Altman2003,Kuklov2003,Duan2003,Powell2009,Hubener2009}, and in the presence of strong repulsions presents finite-temperature phase structure with possibility of checkerboard long range order, supercounterflow, superfluidity, and phase separation \cite{Nakano2012}. Realization of controllable Bose--Bose mixtures \cite{Thalhammer2008} makes way for the experimental reproduction of spin Hamiltonians, with detection and study of magnetic phases such as antiferromagnetic N\'eel and $xy$ ferromagnetic phases.

The physics of coherent coupling in ultracold gases is consequential for quantum information processing and simulation \cite{Gross2017,Lewenstein2012}; supported by the demonstration of strong coupling between bosonic Mott insulators and light \cite{Klinder2015,Landig2016} and the recent work in cold atoms proposing the realization and manipulation of new quantum states \cite{Kollath2016,Mivehvar2017,Sheikhan2019,Schlawin2019a,Schlawin2019b,Curtis2019,Mazza2019,Colella2019}
and high precision quantum limited measurements \cite{Uchino2018}. 
In the two-component Bose gas, introducing coupling reveals fascinating physics; for instance, the phase separation in weak interaction limit \cite{Pethick2001} changing dramatically \cite{Matthews1998,Zibold2010,Nicklas2011,Blakie1999,Search2001,Tommasini2003,Lee2004,Merhasin2005,Abad2013,Butera2017,Chen2017,Shchedrin2018}, modifying the Mott-insulator--superfluid transition \cite{Bornheimer2017}, and altering entanglement properties \cite{Morera2019}. With the introduction of nonlinear coupling in many-body systems, stable collective modes appear, thus paving the way for robust control \cite{Chirikov1979,Buchleitner2002}, with their localization properties and robustness against perturbations demonstrated in a tilted Bose--Hubbard model \cite{Hiller2012}. 

Here, we study Bose gas with two components trapped in a one-dimensional (1D) optical lattice, realizing a two-component Bose--Hubbard model. Presence of multiple bosonic components manifests spin degrees of freedom in addition to Mott insulator and superfluid phases \cite{Kuklov2003,Duan2003}. The coupling of the two components on the nearest-neighbor sites with strong on-site interactions presents unconventional effective ordering generating unprecedented spin behavior. The formation of a new site-dependent non-ferromagnetic spin phase occurs owing to this coupling. The choice of spin alignment along $z$ in this phase can be tuned using hopping parameters. The signature of this unique phase appears as an oscillation in spatial $\zedex$ spin correlation between spins, whereas remaining constant ($\approx 0$) in a conventional anti-ferromagnetic spin phase. In addition to the creation of the unconventional non-ferromagnetic spin phase, the coupling also significantly alters the phase space. When the intra-component interaction is stronger than the inter-component interaction, the introduced coupling switches the phase transition between ferromagnetic and non-ferromagnetic spin phases from first-order to second-order, with the transition width dependent on coupling. When these interactions are comparable, two spin phases (ferromagnetic and non-ferromagnetic) are displayed instead of one (ferromagnetic) with a second-order phase transition, when coupling is introduced. This study provides a spin-independent implementation of the optical lattice for trapping the two-component Bose gas with no additional tuning of on-site interaction strengths and instead leveraging coupling as a parameter to switch between the different spin phases. Multiple perspectives---obtained from methods based on mean-field approximation, exact diagonalization, and entangled-plaquette states (EPS)---enhance the understanding of this system; capturing correlations beyond the scope of a mean-field treatment. The physics of filling factor greater than unity where the novel spin correlations persist, and the consideration of complex intra-component tunneling and inter-component coupling are discussed.

This paper is sectioned as follows: Sec.~\ref{sec:themodel} describes the coupled two-component Bose--Hubbard model and its realization. Sec.~\ref{sec:effspin} details the mapped spin-$\sfrac{1}{2}$ model. Sec.~\ref{sec:methods} presents the different numerical methods considered in the study. Sec.~\ref{sec:results} is the multi-perspective study of the spin phases. Sec.~\ref{sec:spinone} extends to discuss complex intra-component tunneling and inter-component coupling as well as the physics of occupancy greater than unity. Finally, Sec.~\ref{sec:summary} summarizes the results and discusses their implications and future avenues.

\section{\label{sec:themodel}The Model}
\begin{figure}
\includegraphics[width=1\linewidth]{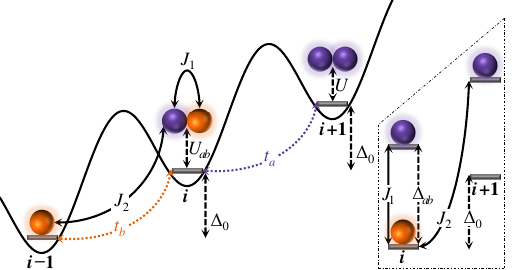}
\caption{\label{fig:ModelN}(color online) Two-component Bose--Hubbard model with on-site inter-($U_{ab}$) and intra-($U$) component interaction ($U \gg U_{ab}$ and $U \sim U_{ab}$) with nearest-neighbor tunneling coefficients $t_a$ and $t_b$. The optical lattice is tilted via external inhomogeneous static electric or magnetic field. \textit{Inset:} To the system is introduced laser-assisted inter-component coupling: $J_1$ (same-site) and $J_2$ (nearest-neighbor). The two components are visualized as two levels of an atom.}
\end{figure}
We consider a 1D system comprising two kinds of bosons trapped in a spin-independent optical lattice as shown in Fig.~\ref{fig:ModelN}. The two components termed as $a$ and $b$, can be envisioned as two internal levels of an atom. We assume low temperature and the optical lattice to be deep enough for the atoms to be confined to the lowest Bloch band describing a two-component Bose--Hubbard model~ \cite{Altman2003,Schachenmayer2015}. The optical lattice is tilted, creating an energy offset between the nearest-neighbor sites as shown in Fig.~\ref{fig:ModelN}, with the assumption that the potential applied to create this tilt is a perturbation. This tilt can be introduced using a magnetic field gradient \cite{Kuno2018,Jaksch2003}. The tilting of the lattice prevents the natural hopping between components $a$ (or $b$) that is obtained via light assisted tunneling: resonant two-photon Raman transition \cite{Kuno2018}. We introduce to this system inter-component coupling; also obtained by a resonant two-photon Raman transition between two internal states as shown in Fig.~\ref{fig:ModelN} (\textit{inset}). The resulting system is governed by the following Hamiltonian: 
\begin{align}
\label{original}
\begin{split}
H = \sum_{i}&(-t_{a} a^+_{i}a_{i+1} -t_{b}b^+_{i}b_{i+1} + H.c.)
\\+&\sum_{k =a,b;i} \dfrac{U}{2}n_{k i}(n_{k i}-1) + \sum_{i} U_{ab}n_{a i}n_{b i} 
\\-&\sum_{i}J_1(a_i^+b_i + b_i^+a_i) - \sum_{i}J_2( b_i^+a_{i+1} + a_{i+1}^+b_{i}),
\end{split}
\end{align}
where $a_{i}$ and $b_{i}$ are the bosonic annihilation operators for components $a$ and $b$ at site $i$, respectively. $t_{k}$ is the component-dependent tunneling parameter. $U$ is the intra- and $U_{ab}$ the inter-component on-site interaction strength. Two kinds of inter-component couplings are present: $J_1$ and $J_2$ representing the same-site and nearest-neighbor coupling, respectively. The energy offset between nearest-neighbor sites of the optical lattice (Fig.~\ref{fig:ModelN}) differentiates the couplings, where $J_2$ defines the inter-component coupling between $b$ at site $i$ and $a$ at site $i+1$. The presence of this offset also allows for tuning of $J_1$ and $J_2$ independently (Fig.~\ref{fig:ModelN} \textit{inset}). In the deep Mott limit ($U,\,U_{ab} \gg t_{k},\,J_1,\,J_2$), with the average number of particles per site being fixed due to the high energy cost of bosons hopping between sites, this model can be mapped to an effective spin system. In the following sections, we study the system in the deep Mott limit, specifically for the case of one particle per site that is mapped to spin-$\sfrac{1}{2}$, where $a$: $\uparrow$ and $b$: $\downarrow$. 

\begin{figure}
\includegraphics{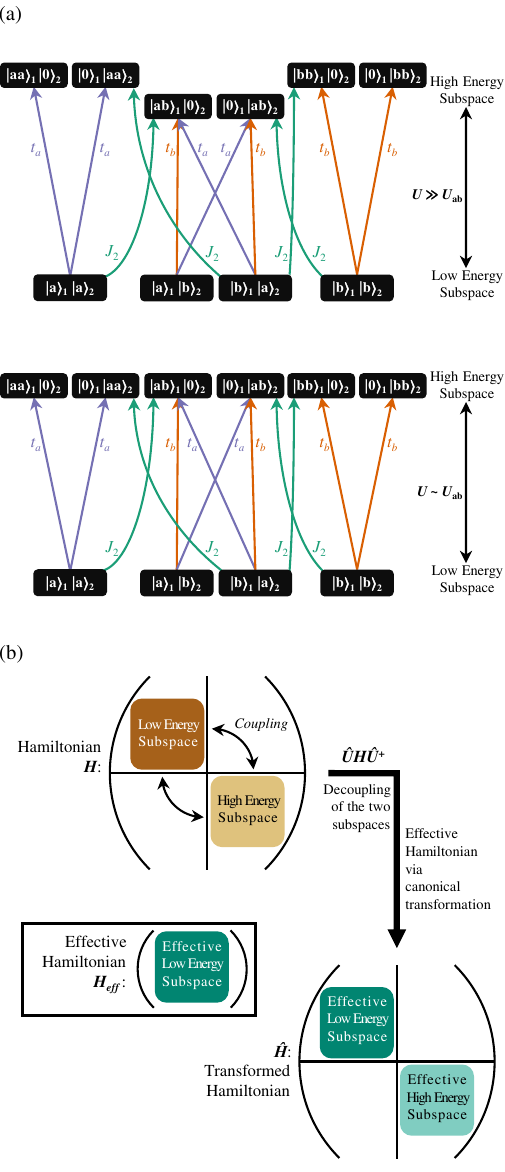}
\caption{\label{fig:subspace}(color online) Schematic of the effective spin Hamiltonian. (a) The low- and high-energy subspaces for a two-site system, roughly separated by the on-site interaction strength and coupled by the inter-component coupling and intra-component tunneling coefficients. (b) The Hamiltonian with low and high-energy subspaces forming the diagonal with the off-diagonal coupling and tunneling coefficients coupling the two subspaces and the effective decoupling of the two subspaces to obtain the effective low-energy spin Hamiltonian.}	
\end{figure}

\section{\label{sec:effspin}Effective Spin Hamiltonian}
Strong on-site interactions divide the Hilbert space into low- and high-energy subspaces, which are coupled by the coupling and tunneling terms as shown in Fig.~\ref{fig:subspace}. A particle hopping from one site to another increases the occupancy, thus going from a low- to high-energy subspace, and thus incurring high energy cost in the process. Whereas, two particles exchanging positions preserve the occupancy and hence do not require a high energy cost. To study the relevant physics, the effective low-energy subspace is obtained by decoupling the existing subspaces; removing any first-order hopping that increases the occupancy and retaining the second-order virtual hopping terms that preserve the occupancy.

The two-component Bose--Hubbard model mapped to the following effective spin (low-energy) Hamiltonian via canonical transformation and followed by a perturbative expansion up to second-order (for details, see Appendix~\ref{sec:appendA}): 
\begin{align}
\label{eff}
\begin{split}
H_{\text{eff}} = \sum_i \Big[&-J_{\perp}(\sigma_i^x\sigma_{i+1}^x + \sigma_i^y\sigma_{i+1}^y) - J_z\sigma_i^z\sigma_{i+1}^z
-h_z(\sigma_i^z) \\&-(h_x+J_1)\sigma_i^x 
+J_{zx}\sigma_i^z\sigma_{i+1}^x - J_{xz}\sigma_i^x\sigma_{i+1}^z\Big] ,
\end{split}
\end{align}
\begin{flalign*}
\text{where }&J_{\perp} = \dfrac{t_at_b}{U_{ab}},
\\& J_z =(t_a^2+t_b^2-J_2^2)\Big(\dfrac{1}{U}-\dfrac{1}{2U_{ab}}\Big),
\\& h_z= \dfrac{2(t_a^2-t_b^2)}{U}, 
\\& h_x = J_2(t_a+t_b)(\dfrac{1}{U_{ab}}+\dfrac{1}{U}), 
\\& J_{zx}= \dfrac{J_2t_b}{U}, \text{ and}
\\& J_{xz} = \dfrac{J_2t_a}{U} &
\end{flalign*}

$J_{\perp}$ and $J_z$ provide the $xy$ and $z$ ordering respectively. The $xy$ ordering is always ferromagnetic in nature, while the $z$ ordering can be tuned from ferromagnetic to anti-ferromagnetic by varying the parameters $t_a$, $t_b$, and $J_2$. The terms $h_z$ and $h_x, J_1$ act as fictitious magnetic fields along $z$ and $x$, respectively, transforming the system to a spin-polarized state. Unconventional ordering terms, $J_{zx}$ and $J_{xz} $ arise due to the directional nearest-neighbor inter-component coupling. They tend to align the spins along $x$ on one site and along $z$ on the nearest-neighbor sites, providing a straightforward implementation for site-dependent spin alignment. That the tunnelings and coupling are light-assisted and the on-site interactions controllable via external magnetic field (Feshbach resonance) allows for the effective parameters to be tuned over a rather large range. Presently only $J_2$ is considered as $J_1$, with no energy cost in same-site coupling, trivially leads the system to an $x$ ferromagnetic phase. With the introduction of nearest-neighbor inter-component coupling, we see new ordering terms $h_x$, $J_{xz},$ and $J_{zx}$; their significance is understood by studying the system behavior dictated by parameters $t_a$, $t_b$, and $J_2$. 

\section{\label{sec:methods}Numerical Methods}
Three different methods are utilized in our study: (a) mean-field approximation, (b) exact diagonalization for small system size ($N<14$), and (c) Variational Monte Carlo with stochastic minimization (VMC-SM) on Entangled-Plaquette States (EPS) for large system size ($N\geq14$). 

Mean-field approximation is the simplest approach pursued. Here, each site is treated independent of the others, making it computationally efficient. Additionally, it is unaffected by the system size. To illustrate the spin phases within mean-field, we designate each alternating site to subspace \textit{A} and its nearest-neighbor sites to subspace \textit{B}. The following variational ansatz is considered:
\begin{align}
\begin{split}
\ket{\Psi} ={}& \prod_{i \in A}(\cos(\theta_A)\ket{a}_i + \sin(\theta_A)\ket{b}_i)
\\& \quad \prod_{i+1 \in B}(\cos(\theta_{B})\ket{a}_{i+1} + \sin(\theta_{B})\ket{b}_{i+1})
\end{split}
\end{align}
Within mean-field approximation, the spin order along $x$ and $z$ computed using Pauli matrices:
\begin{flalign}
\expval{\sigma^x_{i \lor i+1}} = \sin2\theta_{A \lor B}, \quad
\expval{\sigma^z_{i \lor i+1}} = \cos2\theta_{A \lor B}
\end{flalign}
The simplicity of mean-field approximation, in treating each site independent of others, may not accurately reflect the system behavior due to the presence of spin exchanges in the effective spin Hamiltonian correlating nearest-neighbor sites. This could dilute the effect of the unconventional correlations by their being beyond the scope of the mean-field treatment.

Exact diagonalization (ED) provides the most accurate description of the system. Devoid of any approximations, it is the most successful in capturing the effect of the unconventional correlations. Additionally it provides direct comparisons to $N=2$ (see Appendix~\ref{sec:appendB}), where the effective spin Hamiltonian (Eqn.~\ref{eff}) is benchmarked against the original Hamiltonian (Eqn.~\ref{original}) and validated. Although the exact diagonalization does provide a better insight into the system, owing to its large computational expense, this is limited to smaller system sizes $N<14$, where the effect of small system size can be observed. For the thermodynamic limit, arises a need for a numerical method that captures correlations better than the mean-field treatment but with lesser computational expense than exact diagonalization.

Entangled-Plaquette States (EPS) form a class of tensor network states, where the lattice is divided into plaquettes, and the wave function of the system is given by the product of the plaquette wave functions, which are scalar in nature \cite{AlAssam2011}. This goes beyond the mean-field, by considering overlapping or entangled-plaquettes, and is more computationally efficient than exact diagonalization as the number of steps to obtain the system ground state undergo a low-order polynomial increase with system size. It allows for large system sizes while retaining most of the effects of the correlations. In our ansatz with overlapping plaquettes ($P$), each consisting of four sites, coefficients ($C_{P}^{\textbf{n}_P}$; correlator elements) are assigned for all $2^4$ possible configurations in each plaquette. The amplitude for each spin configuration in the state is given as a product of these correlator elements \cite{Mezzacapo2009}.
Variational Monte Carlo (VMC) with stochastic minimization (SM)~\cite{AlAssam2011} is used to obtain the optimized plaquette wave functions that minimize the energy and best describe the ground state. 

Concurrent employment of all three methods in our analysis helps in validating each method and identifying areas where they fail. It also facilitates an accurate interpretation of the results that best reflect the system behavior, with minimal effect of the system size and the approximations made in the mean-field treatment. 

\section{\label{sec:results}Spin Phases}
\begin{figure}
\centering
\includegraphics[width=\linewidth]{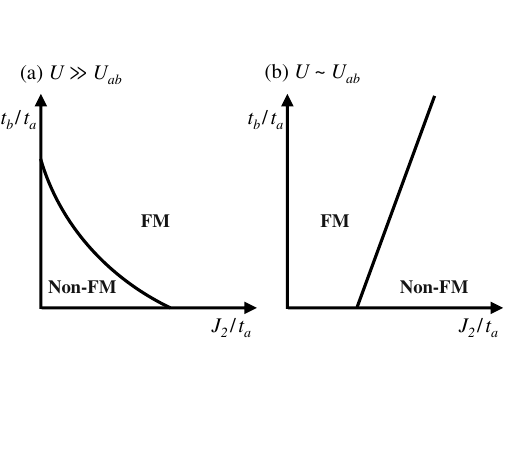}
\caption{\label{fig:SchUggUab}Schematic of spin phases hypothesized to emerge in the presence of inter-component coupling ($J_2$). The phases are shown in the deep Mott regime, (a) $U \gg U_{ab}$ and (b) $U \sim U_{ab}$, and presented as a function of scaled intra-component tunneling $\sfrac{t_b}{t_a}$ and inter-component coupling $\sfrac{J_2}{t_a}$. The FM phase demonstrates typical behavior, with spin along $x$ or $z$. The non-FM phase demonstrates a typical $z$AFM, and an unconventional non-FM phase. A key feature of the unconventional phase: a spin along $z$ on one site correlated with a spin along $x$ on another site.}	
\end{figure}

Previous work studying the two-component Bose--Hubbard model in the absence of inter-component coupling, shows two spin phases: $xy$ ferromagnetic (FM; $t_a\sim t_b$) and $z$ anti-ferromagnetic (AFM; $t_{a \lor b} \gg t_{b \lor a}$) with a first-order phase transition \cite{Altman2003}. This is only seen in the on-site interaction limit $U \gg U_{ab}$. In the other limit $U \sim U_{ab}$, only one spin phase---$z$FM---exists. Motivated by these, we explore the spin phase space in the presence of coupling for strong on-site intra- and inter-component interaction in the limits $U \gg U_{ab}$ and $U \sim U_{ab}$. A schematic of the spin phases in the presence of coupling is shown in Fig.~\ref{fig:SchUggUab}, surmised from the effective Hamiltonian. For $U \gg U_{ab}$: When $t_{a \lor b}\gg t_{b \lor a}$, introducing coupling ($J_2$) evolves the $z$AFM in the non-FM phase to an unconventional non-FM phase arising due to $J_{xz}$ and $J_{zx}$. Further increase in $J_2$ transitions this to $x$FM in the FM phase due to $h_x$, eventually evolving to a $z$FM phase due to $J_z$. When $t_a \sim t_b$ and $J_2 = 0$, an $xy$FM phase is displayed. With coupling, this becomes an $x$FM due to $h_x$. For $U \sim U_{ab}$: with increasing $J_2$, the system evolves from biased $z$FM to $x$FM in the FM phase, transitioning to the unconventional non-FM and evolving to a $z$AFM in the non-FM phase.

\begin{figure*}
\centering
\includegraphics{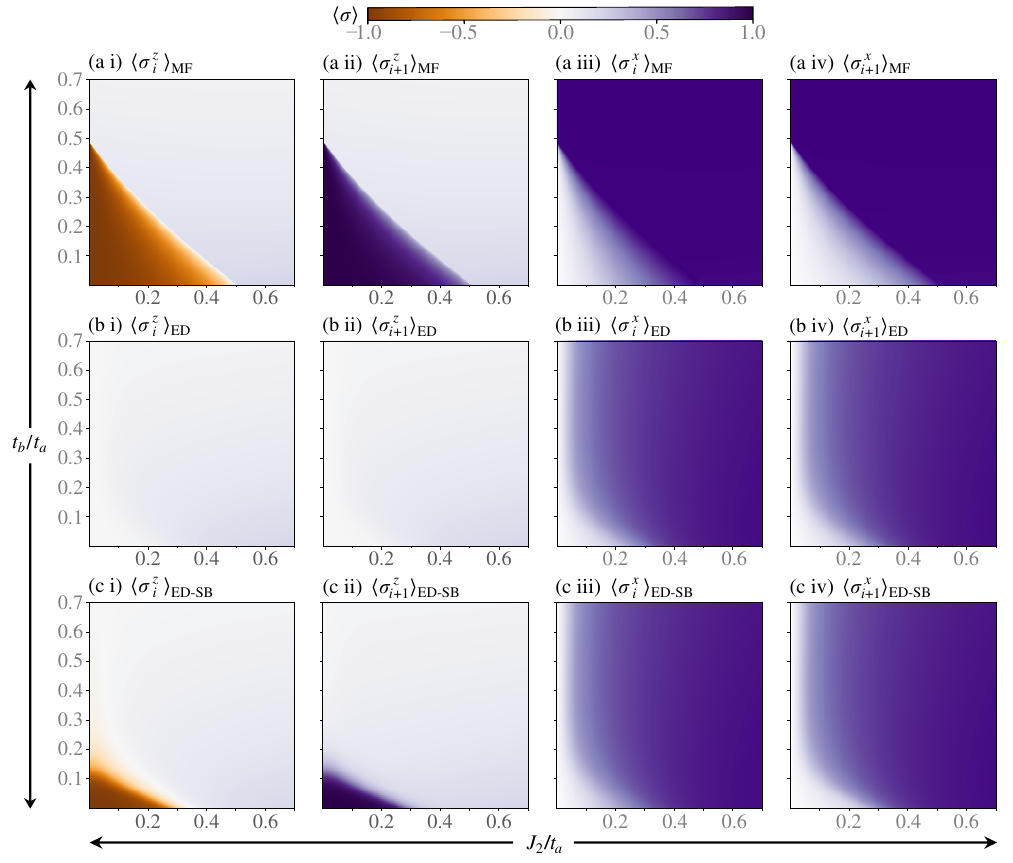}
\caption{\label{fig:order}(color online) Spin order parameters, describing the spin behavior along (i\&ii) $z$ and (iii\&iv) $x$ at site $i$ in subspace $A$ and $i+1$ in $B$. The orders evaluated via (a) mean-field approximation, (b) exact diagonalization for $N=8$ and (c) ED with the symmetry breaking term ($\delta h_z = 4\times10^{-4} t_a$) for $N=8$ are shown as a function of scaled tunneling ($\sfrac{t_b}{t_a}$) and coupling ($\sfrac{J_2}{t_a}$), in the deep Mott regime $(\sfrac{U}{U_{ab}} = 10, \sfrac{U_{ab}}{t_a}=20)$.}	
\end{figure*}

We now introduce coupling to the system and study the spin phases that arise by evaluating the spin order parameters. The site-dependent nature of the unconventional phase emerging due to the unique correlations are confirmed with the examination of the correlation between spins at two different sites. The spin orders along $z$ and $x$ on the nearest-neighbor sites depict the existence of two phases in our system: (1) spins aligning along the same direction (FM) at large tunneling ($\sfrac{t_b}{t_a}$) or coupling ($\sfrac{J_2}{t_a}$) for $U\gg U_{ab}$ (or at tunneling $\geq$ coupling for $U\sim U_{ab}$) and (2) spins aligning along different directions (non-FM) otherwise, as seen in Fig.~\ref{fig:order}. Fig.~\ref{fig:order} (a i \& a ii) representing the spin behavior along $z$ in the mean-field limit, displays a superposition of the quasi-degenerate states (reflecting either of the two degenerate spin configurations---$\uparrow_j\downarrow_{j+1}$ or $\downarrow_j\uparrow_{j+1}$---on the nearest-neighbor sites) in the non-FM phase. Whereas, Fig.~\ref{fig:order} (b i \& b ii) obtained via exact diagonalization with periodic boundary conditions shows the true ground state (reflecting a superposition of the two spin configurations). The ground state obtained via exact diagonalization is symmetric, whereas the state obtained in mean-field treatment has a broken symmetry.

To demonstrate correspondence of the mean-field treatment with exact diagonalization, a weak symmetry breaking term ($-\delta h_z \sum_{i \in (2\mathbb{N}+1)} \sigma_i^z$) is introduced to the effective Hamiltonian and studied via exact diagonalization (ED-SB). The symmetry seen in ED (Fig.~\ref{fig:order} b) is then explicitly broken (Fig.~\ref{fig:order} c). This enforces the selection of one of the two degenerate spin configurations. Spin orders---$\expval{\sigma_i^{z \lor x}}\textsubscript{ED-SB}$---for this symmetry-broken system (Fig.~\ref{fig:order} c i \& c ii) present similar behavior as in the mean-field treatment (Fig.~\ref{fig:order} a i \& a ii). This concordance between ED-SB and the mean-field treatment confirms the latter results in spontaneous symmetry breaking forming a symmetry-broken ground state.

\begin{figure}
\centering
\includegraphics{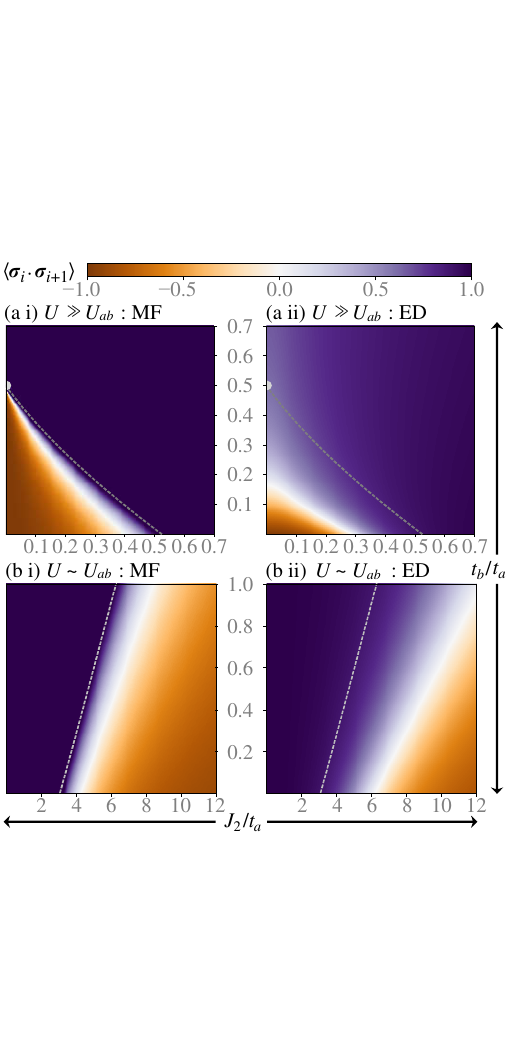}
\caption{\label{fig:PB2}(color online) Spin correlation describing the spin relationship between the nearest-neighbor sites at: (a) $\sfrac{U}{U_{ab}}=10$ and (b) $\sfrac{U}{U_{ab}}=1.2$ evaluated via (i) mean-field approximation and (ii) exact diagonalization for $N=8$. The correlation is shown as a function of scaled tunneling ($\sfrac{t_b}{t_a}$) and coupling ($\sfrac{J_2}{t_a}$), in the deep Mott regime ($\sfrac{U_{ab}}{t_a}=20$). The dashed curve denotes the mean-field critical values $(\sfrac{t_b}{t_a})^C$ from Eqn.~(\ref{cri1}).}
\end{figure}

In the limit $U\gg U_{ab}$ (as well as $U \sim U_{ab}$; see Fig.~\ref{fig:PB2}), the system in the presence of coupling shows a second-order phase transition that appears as a discontinuity in the spin orders. The critical expression for tunneling $\sfrac{t_b}{t_a}$ as a function of coupling $\sfrac{J_2}{t_a}$, and interactions $U,U_{ab}$ describing the phase boundary can be obtained analytically within the mean-field treatment (see Appendix~\ref{sec:appendC}):
\begin{equation}
\label{cri1}
(\sfrac{t_b}{t_a})^{C}_{U\gg U_{ab}}= -A - B \quad \text{and} \quad
(\sfrac{t_b}{t_a})^{C}_{U\sim U_{ab}}= -A + B
\end{equation} 
         where $A = \left(\dfrac{\sfrac{J_2}{t_a}(U+U_{ab})+2U}{2(2U_{ab}-U)}\right)$ and\\
\phantom{where }$B = \sqrt{\left(\dfrac{\sfrac{J_2}{t_a}(U+U_{ab})+2U}{2(2U_{ab}-U)}\right)^2-\left(1-\sfrac{J_2^2}{t_a^2}+\dfrac{\sfrac{J_2}{t_a}(U+U_{ab})}{2U_{ab}-U}\right)}$.\\

The spin correlation ($\expval{\bm{\sigma}_i \cdot \bm{\sigma}_{i+1}}$) between the nearest-neighbor sites is presented in Fig.~\ref{fig:PB2}, which demonstrates the non-FM($<1$) phase and the FM($=1$) phase, with the approximate critical tunneling coefficient $\sfrac{t_b}{t_a}$ describing the phase boundary. For $U \gg U_{ab}$, a first-order transition occurs along the vertical axis (i.e., $J_2=0$). This transitions shifts to second-order as the coupling is introduced, with the transition width increasing with coupling. The spin correlation obtained via exact diagonalization confirms the existence of the two spin phases. The difference between the mean-field approximation and exact diagonalization is expected due to the treatment of all sites independent of each other in the mean-field treatment. Additionally, the exact diagonalization is restricted to smaller system sizes. 

$U\gg U_{ab}:$ In the absence of coupling, dependent on the competing energy parameters $J_z$ and $J_\perp$, the system displays $z$AFM for $J_z\geq J_\perp$ and $xy$FM for $J_z<J_\perp$ \cite{Altman2003}. These phases appear due to the high energy cost of the intermediate state $\ket{aa}_j$ or $\ket{bb}_j$ when compared to $\ket{ab}_j$. This results in a high tunneling probability for states $\ket{a}_j\ket{b}_{j+1}$ or $\ket{b}_j\ket{a}_{j+1}$. 

Introducing coupling to this system, evolves the conventional $z$AFM spin phase to the unconventional non-FM phase as depicted in Fig.~\ref{fig:order}. The unconventional phase demonstrates a spin along $z$ on one site correlated to a spin along $x$ on another site. This results in a stronger $z$ order at one site and $x$ order at the nearest-neighbor sites in the mean-field limit forming a site-dependent spin phase. Spin orientation along $z$ in this phase depends on the intra-component tunneling, with up spin at $\sfrac{t_b}{t_a}<1$ and down spin at $\sfrac{t_b}{t_a}>1$, and thus is intra-component tunneling dependent spin phase. The coupling in the system allows component $b$ on some site $j$ to tunnel to $a$ on site $j+1$ or vice versa. This and the system preference for the high energy intermediate state $\ket{ab}$ allows the nearest-neighbor sites to have the same component ($\ket{a}_j\ket{a}_{j+1}$ or $\ket{b}_j\ket{b}_{j+1}$) and delocalize. These allowed states along with the preferred configuration in the absence of coupling forms the unconventional phase; in the Hamiltonian: emergent due to correlations induced by the $J_{xz}$ and $J_{zx}$ terms and preferentially chosen by the effective ordering $h_x$ and $h_z$. 

The system transitions to the FM phase forming an $x$FM phase upon further increase of the coupling or tunneling. At large tunneling and in the absence of coupling, an $xy$FM phase forms as shown in the previous work \cite{Altman2003}. However, this preference changes with the introduction of coupling as seen in Fig.~\ref{fig:order}, forming an $x$FM phase, due to the strong $h_x$ ordering. 

$U\sim U_{ab}$: Similar spin phases as in $U \gg U_{ab}$ are observed but at different regions of the phase space. In this limit, all the high-energy intermediate states have equivalent energy cost associated with them unlike the previous interaction limit. In a strong coupling limit $(\sfrac{J_2}{t_a} \gg \sfrac{t_a}{t_b},1)$, the conventional $z$AFM spin phase emerges due to the strong effective ordering along $z$: $J_z$ as seen in Fig.~\ref{fig:PB2} (b i \& b ii). With no significant intra-particle tunneling, only the component $b$ at some site $j$ is allowed to tunnel to $a$ at site $j+1$ (or vice versa). The coupling allows for the highest tunneling probability of the configuration $\ket{a}_j\ket{b}_{j+1}$ or $\ket{b}_j\ket{a}_{j+1}$. This results in the formation of the observed $z$AFM phase. Allowing one component to tunnel by a small increase in intra-component tunneling $t_{a \lor b}$ ($J_2 > t_{a \lor b} > t_{b \lor a}$), forms the unconventional non-FM phase.

When one component's intra-component tunneling is as strong as the coupling ($J_2 \sim t_{a \lor b} \gg t_{b \lor a}$), the FM phase exists as seen in Fig.~\ref{fig:PB2} (b i \& b ii). It is expected that when a component is allowed to tunnel, it would be preferred on all sites. This leads to the spin up $z$FM for $t_a \gg t_b$ and spin down $z$FM for $t_b \gg t_a$. However, this preference changes with strong coupling, forming an $x$FM phase due to $h_x$ ordering. With both components being allowed to hop, the system exists in an equal superposition of components. The spin correlations obtained via exact diagonalization are comparable to the mean-field treatment for correlations along $z$ and $x$. 

\begin{figure}
\includegraphics{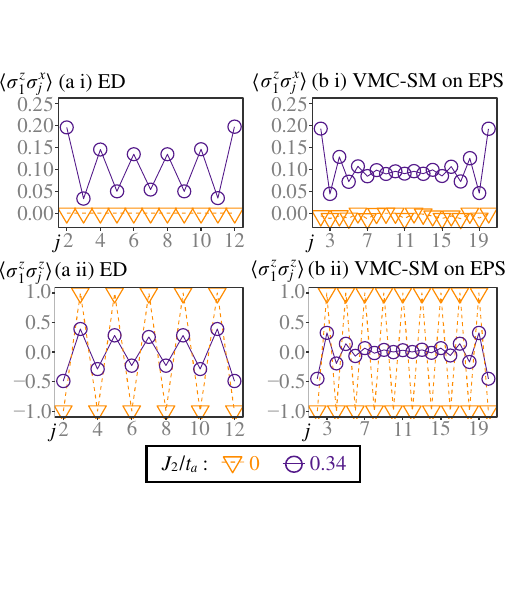}
\caption{\label{fig:sitedep}(color online) Spin correlations describing the spin relationship between site $1$ and some site $j$ along: (i) $\zedex$ and (ii) $z$, for (a) $N=12$ via exact diagonalization and (b) $N=20$ via VMC-SM on EPS. The correlations are shown as a function of site $j$ at $\sfrac{J_2}{t_a} =0.34,\sfrac{t_b}{t_a} = 0.0$ and $\sfrac{J_2}{t_a} \approx 0.0$, $\sfrac{t_b}{t_a} = 0.0$ in the deep Mott regime ($\sfrac{U_{ab}}{t_a}=20$, $\sfrac{U}{U_{ab}}=10$).}	
\end{figure}

\textit{Unconventional non-FM:} The spin orders in exact diagonalization do not display (Fig.~\ref{fig:order}) the conventional $z$AFM and the unconventional non-FM phase as seen in the mean-field treatment. Exact diagonalization shows a superposition of all possible degenerate configurations, hence the spin orders along $z$ show no order in the non-FM region, and the phase can only be identified by the strength of the $x$ order. Whereas, the mean-field treatment shows one of the degenerate configurations as a result of spontaneous symmetry breaking. To demonstrate the existence of the unconventional non-FM phase, the spatial spin correlations are studied along $\zedex$ and $z$ via exact diagonalization ($N=12$) and VMC-SM on EPS ($N=20$). Fig.~\ref{fig:sitedep} (a) presents the evidence of the unconventional site-dependent non-FM phase distinct from the conventional ($z$AFM) spin phase. Illustrated here is the spin correlations between site $1$ and some site \textit{j} along $\zedex$ and $z$ obtained via exact diagonalization for system size $N=12$ at $U\gg U_{ab}$. The unique correlations ($J_{xz}$ and $J_{zx}$ terms in the Hamiltonian) manifests as the observed site-dependent oscillations in the spin correlation along $\zedex$. In the conventional regime, however, the $\zedex$ spin correlation remains constant as a function of site \textit{j}($\approx 0$). The oscillation in the spatial correlation acts as a signature of this unconventional non-FM phase, and an evidence of its existence in exact diagonalization. Similar oscillations appear for $U \sim U_{ab}$ albeit in a large coupling regime ($\sfrac{J_2}{t_a} \approx 5$). Fig.~\ref{fig:sitedep} (b) shows the spin correlations extended to a larger system size of $N=20$ to observe the decay in oscillation of the correlation, expected in 1D owing to the lack of true long-range order. 

\section{\label{sec:spinone}Extension}
In this section we discuss the effect of complex tunneling and coupling coefficients on the spin phases, and extend our study to comment on the effective spin system at occupancy greater than unity. 

\textit{Complex coefficients:} Complex phases are associated with the nearest-neighbor intra-component tunnelings $t_a$ and $t_b$ and the inter-component coupling $J_2$ resulting in new hopping parameters: $t_a = |t_a|e^{i\theta_a}$, $t_b = |t_b|e^{i\theta_b}$, and $J_2=|J_2|e^{i\theta_J}$. The Hamiltonian for the resulting system:
\begin{align}
\begin{split}
H ={}& -\sum_i (|t_a|e^{i\theta_a}a_i^+a_{i+1} + |t_b|e^{i\theta_b}b_i^+b_{i+1} +|J_2|e^{i\theta_J}b_i^+a_{i+1}+H.c.) 
\\&+ \sum_{k =a,b;i} \dfrac{U_{k}}{2}n_{k i}(n_{k i}-1) + \sum_{i} U_{ab}n_{a i}n_{b i} 
\end{split}
\end{align}

The mapped effective spin Hamiltonian in the presence of complex coefficients:
\begin{align}
\begin{split}
H_{\text{eff}} ={}& -J_z\sum_i \sigma_i^z\sigma_{i+1}^z-h_z\sum_i\sigma_i^z
\\& -J_p \sum_i\Big[\cos(\theta_a-\theta_b)(\sigma_i^x\sigma_{i+1}^x+\sigma_i^y\sigma_{i+1}^y)+
\\& \phantom{-J_p \sum_i}\quad\sin(\theta_a-\theta_b)(\sigma_i^x\sigma_{i+1}^y-\sigma_i^y\sigma_{i+1}^x)\Big] 
\\& -h_x\sum_i \Big[(t_a\cos(\theta_a-\theta_J)+t_b\cos(\theta_b-\theta_J))\sigma_i^x-
\\& \phantom{-h_x\sum_i }\quad (t_a\sin(\theta_a-\theta_J)+t_b\sin(\theta_b-\theta_J))\sigma_i^y\Big] 
\\& -J_{xz}\sum_i\cos(\theta_a-\theta_J)\sigma_i^x\sigma_{i+1}^z-\sin(\theta_a-\theta_J)\sigma_i^y\sigma_{i+1}^z
\\& +J_{zx}\sum_i\cos(\theta_b-\theta_J)\sigma_i^z\sigma_{i+1}^x-\sin(\theta_b-\theta_J)\sigma_i^z\sigma_{i+1}^y
\end{split}
\end{align}

The ferromagnetic ordering ($J_p$) of nearest-neighbor sites along $x$ splits to $\exex$ and $\exwhy$ ordering and that along $y$ splits to $\whyex$ and $\whywhy$ ordering. This splitting depends on the phase difference of coefficients $t_a$ and $t_b$. However, the ordering $J_z$ remains unchanged. The fictitious magnetic field along $x$ ($h_x$) splits to $x$ and $y$; this splitting is dependent on the phase difference between tunneling ($t_{a \lor b}$) and coupling ($J_2$). The fictitious magnetic field along $z$ ($h_z$) remains unchanged. A similar effect is seen on the novel correlations with $\exzed$ ordering on nearest-neighbor sites splitting to $\exzed$ and $y\textsf{-}z$ and $\zedex$ splitting to $\zedex$ and $z\textsf{-}y$. These splittings are dependent on the phase difference between $t_{a \lor b}$ and $J_2$. The complex hopping parameters result in an accumulation of phase with each hop. In the case when a component hops to the nearest-neighbor site and back, no net phase is accumulated and consequently the $z$ ordering terms remain unaffected. In any other hopping, a net phase is obtained, which is proportional to the difference of the complex phases associated with tunneling and coupling. Since this is a one-dimensional system, there is only a trivial effect of these complex coefficients; interesting outcomes are expected at higher dimensions.

\textit{Occupancy $>$ 1:} Considering two particles per site, the two-component Bose--Hubbard system is mapped to a spin-$1$ system with basis states $\ket{aa}$, $\ket{ab}$, and $\ket{bb}$ mapped to $\ket{1}$, $\ket{0}$, and $\ket{-1}$, respectively. Considering the limit $U \sim U_{ab}$, since only spin Mott phase ($\ket{0}$) persists for $U \gg U_{ab}$, the effective spin-$1$ Hamiltonian:
\begin{align}
\begin{split}
H_{\text{eff}} ={}& \mu \sum_i (S_i^z)^2 -h^{(1)}_z\sum_i S_i^z -h^{(1)}_x \sum_i S_i^x
\\& -J_z\sum_i S_i^zS_{i+1}^z -J_p \sum_i (S_i^xS_{i+1}^x+S_i^yS_{i+1}^y)
\\& -J_{xz}\sum_i S_i^xS_{i+1}^z +J_{zx} \sum_i S_i^zS_{i+1}^x ,
\end{split}
\end{align}
where $S_i^{\alpha}$ are the spin-$1$ matrices. $\mu = U-U_{ab}$ orders the sites along spin Mott ($\ket{0}$) phase for $\mu >0$ and along $z$ ($\ket{1}$ or $\ket{-1}$) for $\mu <0$. $h^{(1)}_z = \sfrac{6(t_a^2-t_b^2)}{U_{ab}}$ and $h^{(1)}_x = \sfrac{6J_2(t_a+t_b)}{U_{ab}}$ act as fictitious magnetic fields polarizing the spins along $z$ and $x$, respectively. Similar to spin-$\sfrac{1}{2}$, we note $J_p =\sfrac{4t_at_b}{U_{ab}}$ provides ferromagnetic ordering along $x$ and $y$ and $J_z =4\sfrac{(t_a^2+t_b^2-J_2^2)}{U_{ab}}$ provides anti-ferromagnetic or ferromagnetic ordering along $z$ (dependent on the tunneling and coupling parameters $t_a$, $t_b$, and $J_2$). $J_{xz}=\sfrac{4J_2t_a}{U_{ab}}$ and $J_{zx}=\sfrac{4J_2t_b}{U_{ab}}$ provide the novel ordering along $\exzed$ and $\zedex$ on nearest-neighbor sites. As expected, despite increasing the occupancy, the novel ordering persists. In the spin-$1$ system, an additional competing parameter $\mu$ appears with an additional spin Mott phase. 

Generalizing to spin $S=\sfrac{M}{2}$ for occupancy $M$, the effective spin-$S$ Hamiltonian ($S_i^{\alpha}$ are spin $S$ matrices) (Appendix~\ref{sec:appendD}): 
\begin{align}
\begin{split}
H_{\text{eff}} ={}& \mu \sum_i (S_i^z)^2 -h^{(S)}_z\sum_i S_i^z -h^{(S)}_x \sum_i S_i^x
\\& -J_z\sum_i S_i^zS_{i+1}^z -J_p \sum_i (S_i^xS_{i+1}^x+S_i^yS_{i+1}^y)
\\& -J_{xz}\sum_i S_i^xS_{i+1}^z +J_{zx} \sum_i S_i^zS_{i+1}^x ,
\end{split}
\end{align}
where $h^{(S)}_z = 2(2S+1)\dfrac{(t_a^2-t_b^2)}{U_{ab}}$ and \\\phantom{where }$h^{(S)}_x=2(2S+1) \dfrac{J_2(t_a+t_b)}{U_{ab}}$.

\section{\label{sec:summary}Summary}
In conclusion, coupling of the two components in the nearest-neighbor sites via a resonant two-photon Raman transition gives rise to new effective spin Hamiltonian and a unique site-dependent non-FM spin phase. The signature of this unique phase appears as oscillation in spatial $\zedex$ correlation between spins at site $1$ and some site $j$, whereas it remains constant ($\approx 0$) in a conventional $z$AFM spin phase. The phase space notably changes with the introduction of coupling, changing the previously seen first-order phase transition between the FM and non-FM ($z$AFM) spin phases to a second-order phase transition for $U \gg U_{ab}$. The corresponding transition width increases with $\sfrac{J_2}{t_a}$. For $U \sim U_{ab}$, the system goes from having only one (FM) spin phase to two phases (FM and non-FM) with a second-order phase transition between the two. Our analysis demonstrates that coupling of the components in the nearest-neighbor sites provides an easily tunable parameter to switch between the previously seen $z$AFM, $z$FM, and $x$FM spin phases allowing for the implementation of a spin independent optical lattice.

We employ exact diagonalization for small system size ($N < 14$) and Variational Monte Carlo with stochastic minimization on Entangled-Plaquette State for large system size ($N \geq 14$). These confirm the presence of strong unconventional correlations beyond the mean-field approximation. Introduction of complex intra-component tunneling and inter-component coupling alters the ordering along $x$. The unconventional correlations split to have $x$ and $y$ spin orders on a site and $z$ on its nearest-neighbor sites. For filling factor greater than unity, we present the effective spin Hamiltonian in the presence of coupling and demonstrate that the novel correlations persist, and also granting the possibility of transitioning between the different spin phases via the coupling instead of the on-site interactions.

\textit{Future Avenues:} Our work provides new insights into the Bose--Hubbard model and offers an easily tunable phase parameter. The system with coupling can be investigated for symmetry protected topological (SPT) phases \cite{DeLsleuc2019}. A possible extension to two tilted 1D optical lattices parallel to each other can obtain the Su--Schrieffer--Heeger (SSH) model and develop a SPT phase by tuning the coupling and tunneling parameters. Going to higher dimensions, new rich physics is expected \cite{Aidelsburger2013}. Our assessment of the mapped system with complex hopping coefficients leads to the system acquiring a phase with second-order hopping. This is extensible to higher dimensions, where the phase acquired over hopping can have far more interesting consequences: development of Harper's Hamiltonian \cite{Jaksch2003} and interesting spin correlations in the mapped effective spin Hamiltonian. Additionally, the effect of the unconventional correlations for a generalized $M$ occupancy at even and odd fillings using the mapped spin-$\sfrac{M}{2}$ Hamiltonian can be studied. In the absence of coupling, a spin-$1$ system mapped from two-component Bose--Hubbard model with occupancy two is used to lower entropy and obtain ultra-cold temperature \cite{Schachenmayer2015}; this motivates the consideration of occupancy greater than one. The tunability of the system via coupling can provide additional control over the system at higher occupancy. Rydberg atoms have garnered immense relevance for demonstrating interesting phases and system behavior with their long-range van der Waals interactions leading to blockade \cite{Urban2009}, and the resonant dipole--dipole interaction \cite{DeLsleuc2019}. Finally, unexpected spin behaviors may emerge when incorporating Rydberg atoms on account of their interesting properties. 

\begin{acknowledgments}
This study was supported by the National Science Foundation under grant PHY-1912068 and the Welch Foundation under Grant No.~C-1669.
\end{acknowledgments}

\appendix
\section{\label{sec:appendA}Effective Hamiltonian: Canonical Transformation}
Define two complementary subspaces: $H_\mathcal{P}$ (low-energy) with projector $\mathcal{P}$ and $H_\mathcal{Q}$ (high-energy) with projector $\mathcal{Q}(=1-\mathcal{P})$. The effective Hamiltonian (up to second-order) can be expressed as \cite{Cazalilla2003}:
\begin{equation}
H_{\text{eff}} = \mathcal{P}H\mathcal{P}-\mathcal{P}H\mathcal{Q}\dfrac{1}{\mathcal{Q}H\mathcal{Q}}\mathcal{Q}H\mathcal{P} \label{eqn:A1}
\end{equation}
The zeroth-order has no contribution as the hopping terms in the Hamiltonian lead to higher occupancy and are thus projected out. The interaction terms yield $0$ for unit occupancy. In $\mathcal{P}H\mathcal{Q}$ or $\mathcal{Q}H\mathcal{P}$ the only contribution is from hopping terms ($t_a, t_b, J_2$) in the Hamiltonian that can couple the two subspaces. Whereas, in $\mathcal{Q}H\mathcal{Q}$, only the interaction terms ($U, U_{ab}$) contribute as they do not change the subspace.\\
The second-order term $\mathcal{P}H\mathcal{Q}\tfrac{1}{\mathcal{Q}H\mathcal{Q}}\mathcal{Q}H\mathcal{P}$ is:
\begin{align}
\begin{split}
={}& \mathcal{P}\Bigg(\sum_{i}(-t_a a^+_{i}a_{i+1} -t_b b^+_{i}b_{i+1}+ h.c.) - J_2( b_i^+a_{i+1} + a_{i+1}^+b_{i})\Bigg)\mathcal{Q} \\&\left[1/\Bigg(\mathcal{Q}(\sum_{k =a,b;i} \dfrac{U_{k}}{2}n_{k i}(n_{k i}-1) + \sum_{i} U_{ab}n_{a i}n_{b i})\mathcal{Q}\Bigg)\right] \\&\mathcal{Q}\Bigg(\sum_{i}(-t_a a^+_{i}a_{i+1} -t_b b^+_{i}b_{i+1}+ h.c.) - J_2( b_i^+a_{i+1} + a_{i+1}^+b_{i})\Bigg)\mathcal{P}
\end{split}
\end{align}
There are nine second-order or virtual hopping processes that contribute to and form the effective Hamiltonian. The first and second describe two particles of the same component exchanging positions. The eighth and ninth are when component $b$ (or $a$) hops from $(i+1) \lor (i-1)$ to $i$ and component $a$ (or $b$) hops from $i$ to $(i+1) \lor (i-1)$. These four processes have been studied previously~\cite{Altman2003} and result in $xy$FM and either $z$AFM or FM phases. With the introduction of $J_2$, five additional processes arise. The third process describes component $a$ at $i+1$ and $b$ at $i$ exchanging positions. The fourth and sixth together represent $(a \lor b)_i,a_{i+1} \rightarrow b_i,(a \lor b)_{i+1}$. Similarly, the fifth and seventh together describe $b_i,(a \lor b)_{i+1} \rightarrow (a \lor b)_i,a_{i+1}$. These processes are derived and expressed as follows:

\begin{flalign*}
\text{\textcircled{\scriptsize{1}}}\;{}&\dfrac{1}{\mathcal{Q}H\mathcal{Q}} \mathcal{P}\sum_{i}-t_a\left(a^+_{i+1}+a^+_{i-1}\right)a_{i}\mathcal{Q} \sum_{i}-t_a a^+_{i}\left(a_{i+1}+a_{i-1}\right)\mathcal{P} &\\ 
&=\dfrac{t_a^2}{U_{ab}}\sum_{i}\mathcal{P}\left(n_{a,i+1}+n_{a,i-1}\right)\left(a_{i}n_{b,i}a^+_{i}\right)\mathcal{P}\\& \quad +\dfrac{t_a^2}{U}\sum_{i}\mathcal{P}\left(n_{a,i+1}+n_{a,i-1}\right)\left(a_{i}\left(1-n_{b,i}\right)a^+_{i}\right)\mathcal{P} &\\
&=\dfrac{t_a^2}{U_{ab}}\sum_{i}\left(\dfrac{I+\sigma_{i+1}^z}{2}+\dfrac{I+\sigma_{i-1}^z}{2}\right)\left(\dfrac{I-\sigma_{i}^z}{2}\right) \\& \quad +\dfrac{t_a^2}{U}\sum_{i}\left(\dfrac{I+\sigma_{i+1}^z}{2}+\dfrac{I+\sigma_{i-1}^z}{2}\right)\left(I+\sigma_{i}^z\right) \\&
=\dfrac{t_a^2}{2U_{ab}}\sum_{i}\left(I-\sigma_{i}^z\sigma_{i+1}^z\right) +\dfrac{t_a^2}{U}\sum_{i}\left(I+2\sigma_{i}^z +\sigma_{i}^z\sigma_{i+1}^z\right) &
\end{flalign*}
\begin{flalign*}
\text{\textcircled{\scriptsize{2}}}\;{}&\dfrac{1}{\mathcal{Q}H\mathcal{Q}} \mathcal{P}\sum_{i}-t_b\left(b^+_{i+1}+b^+_{i-1}\right)b_{i}\mathcal{Q} \sum_{i}-t_b b^+_{i}\left(b_{i+1}+b_{i-1}\right)\mathcal{P} &\\
&=\dfrac{t_b^2}{2U_{ab}}\sum_{i}\left(I-\sigma_{i}^z\sigma_{i+1}^z\right) +\dfrac{t_b^2}{U}\sum_{i}\left(I-2\sigma_{i}^z +\sigma_{i}^z\sigma_{i+1}^z\right) &\\
\text{\textcircled{\scriptsize{3}}}\;{}&\dfrac{1}{\mathcal{Q}H\mathcal{Q}} \mathcal{P} \sum_{i}-J_2\left(a^+_{i+1}b_{i}+b^+_{i-1}a_{i}\right)\mathcal{Q} \sum_{i}-J_2\left(b^+_{i}a_{i+1}+a^+_{i}b_{i-1}\right)\mathcal{P} &\\
&=\dfrac{J_2^2}{\mathcal{Q}H\mathcal{Q}} \sum_{i} \mathcal{P}\left(n_{a,i+1}b_{i}\mathcal{Q}b^+_{i} + n_{b,i-1}a_{i}\mathcal{Q}a^+_{i}\right)\mathcal{P} &\\
&= \sum_{i} \bigg[ \dfrac{J_2^2}{U}\mathcal{P}n_{a,i+1}\left(1+n_{b,i}\right)\left(1-n_{a,i}\right) + n_{b,i-1}\left(1+n_{a,i}\right)\left(1-n_{b,i}\right)\mathcal{P} + 
\\&\phantom{\sum_{i} \bigg[ \dfrac{J_2^2}{U}\quad}\dfrac{J_2^2}{U_{ab}} \mathcal{P}n_{a,i+1}\left(1+n_{b,i}\right)n_{a,i} + n_{b,i-1}\left(1+n_{a,i}\right)n_{b,i}\mathcal{P} \bigg] &\\
&= \dfrac{J_2^2}{U}\sum_{i}\left(I-\sigma_{i}^z\sigma_{i+1}^z\right) + \dfrac{J_2^2}{2U_{ab}}\sum_{i}\left(I+\sigma_{i}^z\sigma_{i+1}^z\right) &
\end{flalign*}
The fourth and fifth processes together contribute to the fictitious magnetic field $h_x$ and novel correlation $J_{xz}$.
\begin{flalign*}
\text{\textcircled{\scriptsize{4}}}\;{}&\dfrac{1}{\mathcal{Q}H\mathcal{Q}} \mathcal{P}\sum_{i}-t_a\left(a^+_{i+1}+a^+_{i-1}\right)a_{i}\mathcal{Q} \sum_{i}-J_2\left(b^+_{i}a_{i+1}+a^+_{i}b_{i-1}\right)\mathcal{P} &\\
&=\dfrac{J_2t_a}{U_{ab}} \sum_{i} \mathcal{P}\left(n_{a,i+1}b^+_{i}a_{i} + a^+_{i-1}b_{i-1}a_{i}n_{b,i}a^+_{i}\right)\mathcal{P} \\& \quad + \dfrac{J_2t_a}{U} \sum_{i} \mathcal{P} a^+_{i-1}b_{i-1}a_{i}\left(1-n_{b,i}\right)a^+_{i}\mathcal{P} &\\
&=\dfrac{J_2t_a}{2U_{ab}} \sum_{i}\left(\sigma_{i}^x-i\sigma_{i}^y\sigma_{i+1}^z\right)
\\& \quad + \dfrac{J_2t_a}{2U} \sum_{i} \left(\sigma_{i-1}^x+i\sigma_{i-1}^y + \sigma_{i}^z\sigma_{i-1}^x+i\sigma_{i}^z\sigma_{i-1}^y\right) &\\
\text{\textcircled{\scriptsize{5}}}\;{}&\dfrac{1}{\mathcal{Q}H\mathcal{Q}} \mathcal{P}\sum_{i}-J_2\left(a^+_{i+1}b_{i}+b^+_{i-1}a_{i}\right)\mathcal{Q} \sum_{i}-t_a a^+_{i}\left(a_{i+1}+a_{i-1}\right)\mathcal{P} &\\
&=\dfrac{J_2t_a}{\mathcal{Q}H\mathcal{Q}} \sum_{i} \mathcal{P}\left(n_{a,i+1}b_{i}\mathcal{Q}a^+_{i} + b^+_{i-1}a_{i-1}a_{i}\mathcal{Q}a^+_{i}\right)\mathcal{P} &\\
&=\dfrac{J_2t_a}{2U_{ab}} \sum_{i}\left(\sigma_{i}^x+i\sigma_{i}^y\sigma_{i+1}^z\right)
\\& \quad + \dfrac{J_2t_a}{2U} \sum_{i} \left(\sigma_{i-1}^x-i\sigma_{i-1}^y + \sigma_{i}^z\sigma_{i-1}^x-i\sigma_{i}^z\sigma_{i-1}^y\right) &
\end{flalign*}
The sixth and seventh processes together contribute to the fictitious magnetic field $h_x$ and novel correlation $J_{zx}$.
\begin{flalign*}
\text{\textcircled{\scriptsize{6}}}\;{}&\dfrac{1}{\mathcal{Q}H\mathcal{Q}} \mathcal{P}\sum_{i}-t_b\left(b^+_{i+1}+b^+_{i-1}\right)b_{i}\mathcal{Q} \sum_{i}-J_2\left(b^+_{i}a_{i+1}+a^+_{i}b_{i-1}\right)\mathcal{P} &\\
&=\dfrac{J_2t_b}{\mathcal{Q}H\mathcal{Q}} \sum_{i} \mathcal{P}\left(n_{b,i-1}b_{i}\mathcal{Q}a^+_{i} + b^+_{i+1}a_{i+1}b_{i}\mathcal{Q}b^+_{i}\right)\mathcal{P} &\\
&=\dfrac{J_2t_b}{2U_{ab}} \sum_{i}\left(\sigma_{i}^x-i\sigma_{i}^y\sigma_{i-1}^z\right)
\\& \quad + \dfrac{J_2t_b}{2U} \sum_{i} \sigma_{i+1}^x-i\sigma_{i+1}^y - \sigma_{i}^z\sigma_{i+1}^x+i\sigma_{i}^z\sigma_{i+1}^y &
\end{flalign*}

\begin{flalign*}
{}&\text{\textcircled{\scriptsize{7}}}\;\dfrac{1}{\mathcal{Q}H\mathcal{Q}} \mathcal{P}\sum_{i}-J_2\left(a^+_{i+1}b_{i}+b^+_{i-1}a_{i}\right)\mathcal{Q} \sum_{i}-t_b b^+_{i}\left(b_{i+1}+b_{i-1}\right)\mathcal{P} &\\
&=\dfrac{J_2t_b}{\mathcal{Q}H\mathcal{Q}} \sum_{i} \mathcal{P}\left(n_{b,i-1}a_{i}\mathcal{Q}b^+_{i} + a^+_{i+1}b_{i+1}b_{i}\mathcal{Q}b^+_{i}\right)\mathcal{P} &\\
&=\dfrac{J_2t_b}{2U_{ab}} \sum_{i}\left(\sigma_{i}^x+i\sigma_{i}^y\sigma_{i-1}^z\right)
\\& \quad + \dfrac{J_2t_b}{2U} \sum_{i} \sigma_{i+1}^x+i\sigma_{i+1}^y - \sigma_{i}^z\sigma_{i+1}^x-i\sigma_{i}^z\sigma_{i+1}^y &\\
{}&\text{\textcircled{\scriptsize{8}}}\;\dfrac{1}{\mathcal{Q}H\mathcal{Q}} \mathcal{P}\sum_{i}-t_a\left(a^+_{i+1}+a^+_{i-1}\right)a_{i}\mathcal{Q} \sum_{i}-t_b b^+_{i}\left(b_{i+1}+b_{i-1}\right)\mathcal{P} &\\
&=\dfrac{t_at_b}{4U_{ab}} \sum_{i} \bigg[ \left(\sigma_{i+1}^x+i\sigma_{i+1}^y\right)\left(\sigma_{i}^x-i\sigma_{i}^y\right) 
+ \left(\sigma_{i-1}^x+i\sigma_{i-1}^y\right)\left(\sigma_{i}^x-i\sigma_{i}^y\right) \bigg] &\\
&=\dfrac{t_at_b}{2U_{ab}} \sum_{i}\left(\sigma_{i}^x\sigma_{i+1}^x+\sigma_{i}^y\sigma_{i+1}^y\right) &\\
{}&\text{\textcircled{\scriptsize{9}}}\;\dfrac{1}{\mathcal{Q}H\mathcal{Q}} \mathcal{P}\sum_{i}-t_b\left(b^+_{i+1}+b^+_{i-1}\right)b_{i}\mathcal{Q} \sum_{i}-t_a a^+_{i}\left(a_{i+1}+a_{i-1}\right)\mathcal{P} &\\
&=\dfrac{t_at_b}{2U_{ab}} \sum_{i}\left(\sigma_{i}^x\sigma_{i+1}^x+\sigma_{i}^y\sigma_{i+1}^y\right) &
\end{flalign*}
Then, the second term ($\mathcal{P}H\mathcal{Q}\frac{1}{\mathcal{Q}H\mathcal{Q}}\mathcal{Q}H\mathcal{P}$) in the expansion \ref{eqn:A1}:
\begin{align}
\begin{split}
={}& \dfrac{t_at_b}{U_{ab}} \sum_{i}\left(\sigma_{i}^x\sigma_{i+1}^x+\sigma_{i}^y\sigma_{i+1}^y\right) + \left(t_a^2+t_b^2-J_2^2\right)\left(\tfrac{1}{U}-\tfrac{1}{2U_{ab}}\right)\sum_{i} \sigma_{i}^z\sigma_{i+1}^z \\&
+2\dfrac{t_a^2-t_b^2}{U}\sum_i \sigma_i^z + J_2\left(t_a+t_b\right)\left(\dfrac{1}{U}+\dfrac{1}{U_{ab}}\right)\sum_i \sigma_i^x \\& -\dfrac{J_2t_b}{U}\sum_i \sigma_i^z\sigma_{i+1}^x+\dfrac{J_2t_a}{U}\sum_i \sigma_i^x\sigma_{i+1}^z 
\end{split}
\end{align}
Finally, the effective spin Hamiltonian is:
\begin{align}
\begin{split}
H_{\text{eff}} = \sum_i \Big[&-J_{\perp}\left(\sigma_i^x\sigma_{i+1}^x + \sigma_i^y\sigma_{i+1}^y\right) - J_z\sigma_i^z\sigma_{i+1}^z -h_z\left(\sigma_i^z\right) 
\\&-h_x\left(\sigma_i^x\right) +J_{zx}\sigma_i^z\sigma_{i+1}^x - J_{xz}\sigma_i^x\sigma_{i+1}^z\Big]
\end{split}
\end{align}

\section{\label{sec:appendB}N = 2}
\begin{figure*}
\includegraphics{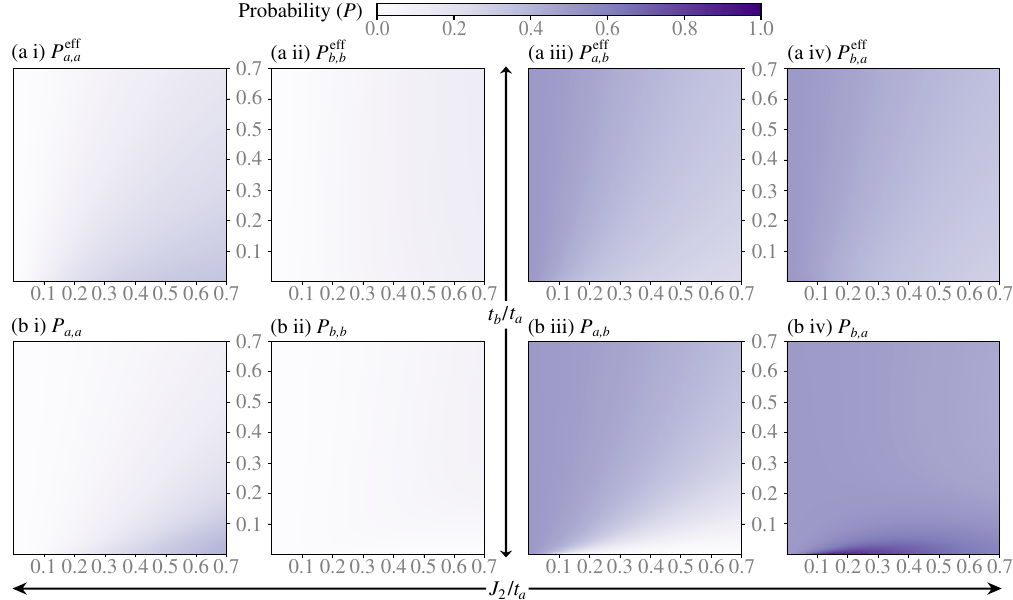}
\caption{\label{fig:N2tatb}(color online) Probability of the basis states of the (a) effective Hamiltonian and (b) low-energy basis states of the original Hamiltonian (i) $\ket{a}_1\ket{a}_2$, (ii) $\ket{b}_1\ket{b}_2$, (iii) $\ket{a}_1\ket{b}_2$, and (iv) $\ket{b}_1\ket{a}_2$. The probabilities are determined in the deep Mott regime ($\sfrac{U_{ab}}{t_a}=20$) $\sfrac{U}{U_{ab}}=10$ and presented as a function of scaled tunneling ($\sfrac{t_b}{t_a}$) and coupling ($\sfrac{J_2}{t_a}$).}	
\end{figure*}
The aim here is to investigate a simplified two-site system as it provides a facile description allowing for the validation of the effective spin system, by comparing the ground states of the original with the mapped two-component Bose--Hubbard Hamiltonian. Consider two sites each hosting component $a \lor b$, forming the basis states $\ket{a}_1\ket{a}_2$,$\ket{a}_1\ket{b}_2$,$\ket{b}_1\ket{a}_2$ and $\ket{b}_1\ket{b}_2$ with the analysis restricted to the limit $U \gg U_{ab}$. The coupling and the tunneling couples these basis states in the original system to high-energy states $\ket{aa}_1\ket{0}_2$, $\ket{0}_1\ket{aa}_2$, $\ket{ab}_1\ket{0}_2$, $\ket{0}_1\ket{ab}_2$, $\ket{bb}_1\ket{0}_2$ and $\ket{0}_1\ket{bb}_2$. The effective spin Hamiltonian and the original Hamiltonian are studied via exact diagonalization with open boundary conditions. In Fig.~\ref{fig:N2tatb} (a), the ground state of the mapped effective spin system is presented as a function of coupling ($\sfrac{J_2}{t_a}$) and tunneling ($\sfrac{t_b}{t_a}$). Upon comparison of the mapped spin system with the ground state of the original system in Fig.~\ref{fig:N2tatb} (b) (low-energy subspace), both systems show a similar ground state composition as a function of tunneling and coupling. At large coupling ($\sfrac{J_2}{t_a} \gg \sfrac{t_b}{t_a}$) the ground state compositions of the two systems deviate, where the mapped system shows a lower probability of the state $\ket{b}_1\ket{a}_2$ and thus higher probability of the other three basis states. At higher values of coupling, the Mott condition required for the mapping fails. This leads to a negligible but non-zero probability of high energy states $\ket{aa}_1\ket{0}_2$, $\ket{0}_1\ket{aa}_2$, $\ket{ab}_1\ket{0}_2$, $\ket{0}_1\ket{ab}_2$, $\ket{bb}_1\ket{0}_2$, and $\ket{0}_1\ket{bb}_2$. Additionally, the mapped effective Hamiltonian works best for a large system size and in the derivation of the effective Hamiltonian the expansion is limited to second-order; inclusion of higher order terms can reduce the deviation seen in the ground state composition between the mapped and original system. Studying the probability of basis states, we confirm that the mapped system is an appropriate representation for the parameter strengths considered. We explain the ground state of mapped system by studying the effective Hamiltonian. The preference for $\ket{b}_1\ket{a}_2$ appears due to $J_z \propto J_2$, and for $\ket{a}_1\ket{a}_2$ or $\ket{b}_1\ket{b}_2$ due to the ordering $J_z, h_z \propto t_{a \lor b}$, whereas the preference for the superposition of states $\ket{a}_1\ket{a}_2$ and $\ket{b}_1\ket{a}_2$ or $\ket{b}_1\ket{b}_2$ and $\ket{b}_1\ket{a}_2$ emerges due to $J_{zx}$ or $J_{xz}$. For the interaction limit $U \sim U_{ab}$, a similar conclusion can be made.

\section{\label{sec:appendC}Analytical Phase Boundary}
Within the mean-field approximation, with the variational ansatz defined, the energy per site:
\begin{align*}
\begin{split}
E_{MF} =& -J_z\cos2\theta_A\cos2\theta_B-J_p\sin2\theta_A\sin2\theta_B \\
& -\dfrac{h_z}{2}(\cos2\theta_A+\cos2\theta_B) -\dfrac{h_x}{2}(\sin2\theta_A+\sin2\theta_B)\\
& -\dfrac{(J_{xz}-J_{zx})}{2}(\cos2\theta_A\sin2\theta_B+\sin2\theta_A\cos2\theta_B)
\end{split}
\end{align*} 
Minimizing energy with respect to $\theta_B$ ($\sfrac{\partial E}{\partial \theta_B}=0$): 
\begin{align*}
\sin2\theta_B =\dfrac{K_1}{\sqrt{K_1^2+K_2^2}}, \quad \cos2\theta_B =\dfrac{K_2}{\sqrt{K_1^2+K_2^2}} ,
\end{align*}
where $K_1 = J_p\sin2\theta_A +\dfrac{(J_{xz}-J_{zx})}{2}\cos2\theta_A+\dfrac{h_x}{2}$ and\\\phantom{where }$K_2 = J_z\cos2\theta_A +\dfrac{(J_{xz}-J_{zx})}{2}\sin2\theta_A+\dfrac{h_z}{2}$.
\\ Energy after eliminating $\theta_B$:
\begin{align*}
E_{MF} = -\dfrac{1}{2}(h_z\cos2\theta_A+h_x\sin2\theta_A)-\sqrt{K_1^2+K_2^2}
\end{align*}
The energy is too complicated to solve for $\theta_A$, approximations are made to obtain analytical expressions for $\theta_A$. 

$U\gg U_{ab}$: We consider two approximations. In approximation $(i)$, consider only $J_p$,$J_z$ and $h_x$. This is valid for small and intermediate $\sfrac{t_b}{t_a}$ and $\sfrac{J_2}{t_a}$. Under this assumption, the non-FM (NFM) phase is approximately derived from the mean-field energy:
\begin{equation*}
E_{MF} = -\dfrac{h_x\sin2\theta_A}{2}-\sqrt{(J_z^2-J_p^2)\cos^22\theta_A+J_ph_x\sin2\theta_A+J_p^2+\dfrac{h_x^2}{4}}
\end{equation*}
Minimizing energy with respect to $\theta_A$ and solving for $\theta_A$:
\begin{align*}
\begin{split}
&\sin2\theta^{\text{NFM},1 \lor 2}_A = \dfrac{J_ph_x \pm |J_z|h_x}{2(J_z^2-J_p^2)}\\&
\sin2\theta^{\text{NFM},3}_A = 1
\end{split}
\end{align*}
$\theta^{\text{NFM},1}_A$ fits the numerically obtained $\theta_A$ best among the three in the non-FM region.

In approximation $(ii)$, consider each of the following cases:
\begin{enumerate}
\item $J_z,h_x$ (large $\sfrac{J_2}{t_a}$ and small $\sfrac{t_b}{t_a}$) or 
\item $J_p,J_z$ (small $\sfrac{J_2}{t_a}$ and large $\sfrac{t_b}{t_a}$) or 
\item $J_p,h_x$ (large $\sfrac{J_2}{t_a}$ and large $\sfrac{t_b}{t_a}$),
\end{enumerate} 
where the terms are significant. In all of the three cases: 
\begin{equation*}
\cos2\theta^{FM}_A = 0
\end{equation*}
The critical value of $\sfrac{t_b}{t_a}$, going from non-FM to FM, obtained by finding the boundary between the two analytical $\theta_A$ is computed: 
\begin{align*}
\begin{split}
(\sfrac{t_b}{t_a})^{C}_{U\gg U_{ab}}=& -\left(\dfrac{\sfrac{J_2}{t_a}(U+U_{ab})+2U}{2(2U_{ab}-U)}\right)-\\& \sqrt{\left(\dfrac{\sfrac{J_2}{t_a}(U+U_{ab})+2U}{2(2U_{ab}-U)}\right)^2-\left(1-\sfrac{J_2^2}{t_a^2}+\dfrac{\sfrac{J_2}{t_a}(U+U_{ab})}{2U_{ab}-U}\right)}
\end{split}
\end{align*} 
$U\sim U_{ab}$: We consider two approximations. In approximation $(i)$, set $h_z$ and $J_{xz}-J_{zx}$ to 0. This is valid when the intra-component tunnelings are small ($t_b,t_a \ll J_2$), or comparable to each other ($t_a \sim t_b$). This limit is able to approximately capture the non-FM phase from the mean-field energy:
\begin{equation*}
E_{MF} = -\dfrac{h_x\sin2\theta_A}{2}-\sqrt{(J_z^2-J_p^2)\cos^22\theta_A+J_ph_x\sin2\theta_A+J_p^2+\dfrac{h_x^2}{4}}
\end{equation*}
Minimizing energy with respect to $\theta_A$ and solving for $\theta_A$:
\begin{align*}
\begin{split}
&\sin2\theta^{\text{NFM},1 \lor 2}_A = \dfrac{J_ph_x \pm |J_z|h_x}{2(J_z^2-J_p^2)}\\&
\sin2\theta^{\text{NFM},3}_A = 1
\end{split}
\end{align*}
$\theta^{\text{NFM},1}_A$ fits the numerically obtained $\theta_A$ best among the three in the non-FM region. 

In approximation $(ii)$, set $J_z$, $J_p$ and $J_{xz}-J_{zx}$ to 0. This solution set only considers the spin polarizing terms $h_z$ and $h_x$, and is able to capture the FM phase. This approximation is valid for small coupling $\sfrac{J_2}{t_a}$. It however fails to capture the region around $\sfrac{t_b}{t_a}\sim 1$ and $\sfrac{J_2}{t_a}\rightarrow 0$, where both the ordering terms considered would tend to $0$.
\begin{equation*}
E_{MF} = -\dfrac{h_z\cos2\theta_A}{2}-\dfrac{h_x\sin2\theta_A}{2}-\sqrt{\dfrac{h_z^2+h_x^2}{4}}
\end{equation*}
Minimizing energy with respect to $\theta_A$ and solving for $\theta_A$
\begin{equation*}
\cos2\theta^{FM}_A = \dfrac{h_z}{\sqrt{h_z^2+h_x^2}}
\end{equation*}
The critical value of $\sfrac{t_b}{t_a}$, going from non-FM to FM, obtained by finding the boundary between the two analytical $\theta_A$ is computed ($\theta^{\text{NFM},1}_A = lim_{h_z\rightarrow 0}\theta^{FM}_A =\sfrac{\pi}{4}$):
\begin{align*}
\begin{split}
(\sfrac{t_b}{t_a})^{C}_{U\sim U_{ab}}=& -\left(\dfrac{\sfrac{J_2}{t_a}(U+U_{ab})+2U}{2(2U_{ab}-U)}\right)+\\& \sqrt{\left(\dfrac{\sfrac{J_2}{t_a}(U+U_{ab})+2U}{2(2U_{ab}-U)}\right)^2-\left(1-\sfrac{J_2^2}{t_a^2}+\dfrac{\sfrac{J_2}{t_a}(U+U_{ab})}{2U_{ab}-U}\right)}
\end{split}
\end{align*} 

\section{\label{sec:appendD}Mapping M particles per site}
Generalizing the system to $M$ particles per site, with the analysis limited to $U-U_{ab} = \mu \ll U,U_{ab}$, consider a site having $n_a$ particles of component $a$ and $n_b$ of component $b$, such that $n_a+n_b=M$. To map the two-component Bose--Hubbard model to an effective low-energy Hamiltonian, define the subspaces $P$ (occupancy = $M$) and $Q$ (occupancy > $M$).\\~\\
\noindent The effective Hamiltonian up to second-order: 
\begin{equation*}
H_{\text{eff}} = PHP-PHQ\dfrac{1}{QHQ-PHP}QHP
\end{equation*}
Mapping the bosonic creation, annihilation, and number operators to Spin-$\sfrac{M}{2}$ matrices:
\begin{align*}
\begin{split}
& n_{a,i} = \dfrac{M}{2}+S_i^z, \quad n_{b,i} = \dfrac{M}{2}-S_i^z \\&
a_i^+b_i = S_i^x+iS_i^y, \quad b_i^+a_i = S_i^x-iS_i^y
\end{split}
\end{align*}
The first term in the Hamiltonian:
\begin{align*}
\begin{split}
PHP &= \dfrac{U}{2}n_{a,i}(n_{a,i}-1) +\dfrac{U}{2}n_{b,i}(n_{b,i}-1) +U_{ab}n_{a,i}n_{b,i} \\&
=\dfrac{U}{2}M(M-1)-\mu\dfrac{M^2}{4}+\mu(S_i^z)^2
\end{split}
\end{align*}
The second term in the Hamiltonian, $-PHQ\tfrac{1}{QHQ-PHP}QHP$, is split into 4 terms:
\begin{flalign*}
(a)~ & -t_a^2P\left[(n_{a,i+1}+n_{a,i-1})a_i\tfrac{Q}{QHQ-PHP}a_i^+\right]P&&\\
& -t_b^2P\left[(n_{b,i+1}+n_{b,i-1})b_i\tfrac{Q}{QHQ-PHP}b_i^+\right]P&&\\
& -J_2^2P\left[n_{a,i+1}b_i\tfrac{Q}{QHQ-PHP}b_i^++n_{b,i-1}a_i\tfrac{Q}{QHQ-PHP}a_i^+\right]P&&\\
(b)~ & -t_at_bP\Big[(a_{i+1}^+b_{i+1}+a_{i-1}^+b_{i-1})a_i\tfrac{Q}{QHQ-PHP}b_i^+ &&\\
& \phantom{-t_at_bP\Big[}+(b_{i+1}^+a_{i+1}+b_{i-1}^+a_{i-1})b_i\tfrac{Q}{QHQ-PHP}a_i^+ \Big]P&&\\
(c)~ & -J_2t_aP\Big[n_{a,i+1}\Big(a_i\tfrac{Q}{QHQ-PHP}b_i^+ +b_i\tfrac{Q}{QHQ-PHP}a_i^+\Big) &&\\
& \phantom{-J_2t_aP\Bigg[}+ (a_{i-1}^+b_{i-1}+b_{i-1}^+a_{i-1})a_i\tfrac{Q}{QHQ-PHP}a_i^+\Big]P&&\\
(d)~ & -J_2t_bP\Big[n_{b,i-1}\Big(a_i\tfrac{Q}{QHQ-PHP}b_i^+ +b_i\tfrac{Q}{QHQ-PHP}a_i^+\Big) &&\\
& \phantom{-J_2t_bP\Bigg[} + (a_{i+1}^+b_{i+1}+b_{i+1}^+a_{i+1})b_i\tfrac{Q}{QHQ-PHP}b_i^+\Big]P&&
\end{flalign*}
Focusing on $(a)$, $QHQ-PHP \approx U_{ab}M$; $a_iQa_i^+$
\begin{flalign*}
& = \sum_{n_b=0}^M a_i\left[\prod_{k=0,k\neq n_b}^N\dfrac{(k-n_{b,i})}{(k-n_b)}\right]a_i^+ =\sum_{n_b=0}^M (n_a+1)\left[\prod_{k=0,k\neq n_b}^N\dfrac{(k-n_{b,i})}{(k-n_b)}\right]&&\\
& =\sum_{n_b=0}^M (M+1-n_b)\left[\dfrac{1}{n_b!}\dfrac{1}{(M-n_b)!}\right] \left[(-1)^{n_b}\textstyle\prod_{k=0,k\neq n_b}^M (k-n_{b,i})\right]&&\\
& =\dfrac{1}{M!}\sum_{n_b=0}^M (-1)^{n_b} (M+1-n_b) \binom{M}{n_b} \left[\textstyle\prod_{k=0,k\neq n_b}^M (k-n_{b,i})\right]&&
\end{flalign*}
It follows that:
\begin{flalign*}
& b_iQb_i^+ =\dfrac{1}{M!}\sum_{n_a=0}^M (-1)^{n_a} (M+1-n_a) \binom{M}{n_a} \left[\textstyle\prod_{k=0,k\neq n_a}^M (k-n_{a,i})\right]&&
\end{flalign*}

Simplification of the terms in the above expression: 
\begin{flalign*}
& \prod_{k=0,k\neq n_b}^M (k-n_{b,i}) = C_0 + C_1n_{b,i} +C_2n_{b,i}^2 + \ldots + C_Mn_{b,i}^M &&\\
&\qquad C_M = (-1)^M&&\\
&\qquad C_{M-1}=(-1)^Mn_b+(-1)^{M-1}(\textstyle\sum_{n_1=0}^Mn_1), &&\\
&\qquad C_{M-2} = (-1)^Mn_b^2 +(-1)^{M-1}(\textstyle\sum_{n_1=0}^Mn_1)n_b &&\\
&\qquad \phantom{C_{M-2} =} +(-1)^{M-2}(\textstyle\sum_{\substack{n_1=0\\n_2>n_1}}^Mn_1n_2) + \ldots &&\\
&\qquad C_1 = (-1)^Mn_b^{M-1}+\ldots,&&\\
&\qquad C_0 = (-1)^Mn_b^M +(-1)^{M-1}(\textstyle\sum_{n_1=0}^Mn_1)n_b^{M-1}+\ldots&&\\
&\qquad C_{M-k} = (-1)^Mn_b^k +(-1)^{M-1}(\textstyle\sum_{n_1=0}^Mn_1)n_b^{k-1}+\ldots&&\\
& \sum_{n_b}(-1)^{n_b} n_b^k \binom{M}{n_b} =
\begin{dcases}
(-1)^MM!, &\text{if } k = M\\
       0, &           k < M
\end{dcases}&&\\
& n_b^M = n_b(n_b-1)(n_b-2)\ldots(n_b-(M-1)) + (\textstyle\sum_{n_1=0}^{M-1}n_1)n_b^{M-1}&&
\end{flalign*}
Substituting these simplifications: $a_iQa_i^+$
\begin{flalign*}
=&\dfrac{1}{M!}\sum_{n_b=0}^M (-1)^{n_b} (M+1-n_b) \binom{M}{n_b} \left[\prod_{k=0,k\neq n_b}^M (k-n_{b,i})\right]&& \\
=&\dfrac{1}{M!}\sum_{n_b=0}^M (M+1)\left( (-1)^{n_b}\binom{M}{n_b} (-1)^M n_b^M \right)&&\\
& -\dfrac{1}{M!}\sum_{n_b=0}^M n_b (-1)^{n_b}\binom{M}{n_b} \Bigg((-1)^M n_b^M + (-1)^{M-1}\left(\sum_{n=0}^Mn\right)n_b^{M-1} &&\\
& \phantom{-\dfrac{1}{M!}\sum_{n_b=0}^M n_b (-1)^{n_b}\binom{M}{n_b} \Bigg(} + (-1)^M n_b^{M-1}n_{b,i}\Bigg) &&\\
=&(M+1)-n_{b,i} -\dfrac{1}{M!}\sum_{n_b=0}^M (-1)^{n_b}\binom{M}{n_b} \Bigg((-1)^M n_bn_b(n_b-1)\ldots &&\\
&\ldots(n_b-(M-1))+(-1)^M \left(\sum_{n=0}^{M-1}n\right)n_b^M + (-1)^{M-1}\left(\sum_{n=0}^Mn\right)n_b^{M} \Bigg) &&\\
=&(M+1)-n_{b,i} -M\\&-\dfrac{1}{M!}\sum_{n_b=0}^M (-1)^{n_b}\binom{M}{n_b} (-1)^M \left(\dfrac{M(M-1)}{2}-\dfrac{M(M+1)}{2}\right) n_b^M&&\\
=&(M+1)-n_{b,i} -M+M = (M+1)-n_{b,i}&&
\end{flalign*}
\begin{flalign*}
\text{It follows that } b_i&Qb_i^+ = (M+1)-n_{a,i} &&
\end{flalign*}
Simplifying (a) in the expression:\\($z$: number of nearest neighbors)
\begin{flalign*}
(a) ={}& -\dfrac{t_a^2}{MU_{ab}}P\left[(n_{a,i+1}+n_{a,i-1})(M+1-n_{b,i})\right]P&&\\
&-\dfrac{t_b^2}{MU_{ab}}P\left[(n_{b,i+1}+n_{b,i-1})(M+1-n_{a,i})\right]P&&\\
{}&-\dfrac{J_2^2}{MU_{ab}}P\left[n_{b,i+1}(M+1-n_{b,i})+n_{a,i-1}(M+1-n_{a,i})\right]P&& \\
={}&-\dfrac{zt_a^2}{MU_{ab}}\left(\text{Const.} + (M+1)S_i^z+S_i^zS_{i+1}^z\right)&&\\
&-\dfrac{zt_b^2}{MU_{ab}}\left(\text{Const.} - (M+1)S_i^z+S_i^zS_{i+1}^z\right)+\dfrac{zJ_2^2}{MU_{ab}}S_i^zS_{i+1}^z&&
\end{flalign*}
Considering the other terms in the second expression of the Effective Hamiltonian: 
(For (b), (c), and (d), $QHQ-PHP = \sfrac{U}{2}(M-1)+\sfrac{U_{ab}}{2}(M+1)\approx U_{ab}M$):
\begin{flalign*}
(b)={}&-t_at_bP\Bigg[(a_{i+1}^+b_{i+1}+a_{i-1}^+b_{i-1})a_i\tfrac{Q}{QHQ-PHP}b_i^+ &&\\
&\phantom{-t_at_bP\Bigg[} + (b_{i+1}^+a_{i+1}+b_{i-1}^+a_{i-1})b_i\tfrac{Q}{QHQ-PHP}a_i^+\Bigg]P&&\\
={}& -\dfrac{t_at_b}{MU_{ab}}P\Bigg[(a_{i+1}^+b_{i+1}+a_{i-1}^+b_{i-1})a_i\left(\sum_{n_a}\prod_{k=0,k\neq n_a}^M\dfrac{(k-n_{a,i})}{(k-n_a)}\right)b_i^+&&\\
& \qquad + (b_{i+1}^+a_{i+1}+b_{i-1}^+a_{i-1})b_i\left(\sum_{n_b}\prod_{k=0,k\neq n_b}^M\dfrac{(k-n_{b,i})}{(k-n_b)}\right)a_i^+\Bigg]P&&
\end{flalign*}
\begin{flalign*}
={}& -\dfrac{t_at_b}{MU_{ab}}P\Bigg[(a_{i+1}^+b_{i+1}+a_{i-1}^+b_{i-1})b_i^+a_i\dfrac{M!}{M!} &&\\
&\phantom{-\dfrac{t_at_b}{MU_{ab}}P\Bigg[} + (b_{i+1}^+a_{i+1}+b_{i-1}^+a_{i-1})a_i^+b_i\dfrac{M!}{M!}\Bigg]P&&\\
={}& -\dfrac{zt_at_b}{MU_{ab}}\left((S_{i+1}^x+iS_{i+1}^y)(S_{i}^x-iS_{i}^y) + (S_{i+1}^x-iS_{i+1}^y)(S_{i}^x+iS_{i}^y)\right)&&\\
={}& -\dfrac{z2t_at_b}{MU_{ab}}\left(S_{i}^xS_{i+1}^x + S_{i}^yS_{i+1}^y\right)&&\\
(c) ={}& -J_2t_aP\Bigg[n_{a,i+1}\left(a_i\tfrac{Q}{QHQ-PHP}b_i^+ +b_i\tfrac{Q}{QHQ-PHP}a_i^+\right) &&\\
&\phantom{-J_2t_aP\Bigg[}+ (a_{i-1}^+b_{i-1}+b_{i-1}^+a_{i-1})a_i\tfrac{Q}{QHQ-PHP}a_i^+\Bigg]P&&\\
={}& -\dfrac{J_2t_a}{MU_{ab}}(2S_i^x)\left(\dfrac{M}{2}+S_{i+1}^z\right) -\dfrac{J_2t_a}{MU_{ab}}(2S_{i-1}^x)\left(\dfrac{M}{2}+1+S_{i}^z\right)&&\\
={}& -\dfrac{zJ_2t_a}{MU_{ab}}((M+1)S_i^x+2S_i^xS_{i+1}^z)&&\\
(d)={}& -J_2t_bP\Bigg[n_{b,i-1}\left(a_i\tfrac{Q}{QHQ-PHP}b_i^+ +b_i\tfrac{Q}{QHQ-PHP}a_i^+\right)&&\\
&+ (a_{i+1}^+b_{i+1}+b_{i+1}^+a_{i+1})b_i\tfrac{Q}{QHQ-PHP}b_i^+\Bigg]P&&\\
={}& -\dfrac{J_2t_b}{MU_{ab}}(2S_i^x)\left(\dfrac{M}{2}-S_{i-1}^z\right) -\dfrac{J_2t_b}{MU_{ab}}(2S_{i+1}^x)\left(\dfrac{M}{2}+1-S_{i}^z\right)&&\\
={}& -\dfrac{zJ_2t_b}{MU_{ab}}((M+1)S_i^x-2S_i^xS_{i+1}^z)&&
\end{flalign*}

The effective Hamiltonian, scaled by $M(=2S)$:
\begin{widetext}
\begin{align*}
H_{\text{eff}} =& \mu\sum_i(S_i^z)^2 -z(2S+1)\sum_i\dfrac{(t_a^2-t_b^2)}{U_{ab}}S_i^z -z(2S+1)\sum_i\dfrac{J_2(t_a+t_b)}{U_{ab}}S_i^x 
-\dfrac{2(t_a^2+t_b^2-J_2^2)}{U_{ab}}\sum_iS_i^zS_{i+1}^z\\ &-\dfrac{4t_at_b}{U_{ab}}\sum_i(S_i^xS_{i+1}^x+S_i^yS_{i+1}^y) 
-\dfrac{4J_2t_a}{U_{ab}}\sum_iS_i^xS_{i+1}^z+\dfrac{4J_2t_b}{U_{ab}}\sum_iS_i^zS_{i+1}^x
\end{align*}
\end{widetext}

\bibliography{references}
\end{document}